\documentstyle[aps,prl,twocolumn,epsf]{revtex}

\def \imki{Im$\chi({\bf q},\omega)$}

\newcommand{\beq}{\begin{equation}}
\newcommand{\eeq}{\end{equation}}
\newcommand{\beqa}{\begin{eqnarray}}
\newcommand{\eeqa}{\end{eqnarray}}

\renewcommand{\lambda}{\ell}

\newcommand{\bq}{{\bf q}}

\begin{document}
\draft
\twocolumn[\hsize\textwidth\columnwidth\hsize\csname @twocolumnfalse\endcsname
\title{Linear dependence of  peak width in 
$\chi(\bq, \omega)$ vs T$_c$ for YBCO superconductors}
 
\author{  A.V. Balatsky$^{(1)}$  and P. Bourges $^{(2)}$}
\date{\today}

\address{$^{(1)}$ T-Div and MST-Div, Los Alamos National Laboratory, 
Los Alamos, NM 87545, USA}
\address{$^{(2)}$  Laboratoire L\'eon Brillouin, CE Saclay, 
91191 Gif/Yvette ,France}


\maketitle
\begin{abstract}
It is shown that the momentum space width of the  peak in the
spin susceptibility, Im$\chi(\bq,\omega)$, is {\em linearly} proportional 
to the superconducting $T_c$: $T_c = \hbar v^*\Delta q$ with 
$\hbar v^* \simeq 35 meV$\AA.   This relation is similar to the 
linear relation between incommensurate peak splitting and $T_c$ 
in LaSrCuO superconductors, as first proposed by Yamada {\it et al}
($Phys. \ Rev.  \ {\bf B 57}, 6165, (1998)$). The velocity $\hbar v^*$ 
is smaller than Fermi velocity or the spin-wave velocity of the 
parent compound and remains the same for a wide doping range. This result
points towards strong similarities in magnetic state of YBCO and LaSrCuO. 
\end{abstract}
\noindent PACS numbers: 74.20.-z, 78.70.Nx, 61.12.-q 

\
\
]
 
Recent progress in neutron scattering in high-$T_c$ superconductors
system, ${\rm YBa_2Cu_3O_{6+x}}$ (YBCO),
allowed to gather a wide variety of inelastic neutron scattering data
which reveal a nontrivial structure of the antiferromagnetic susceptibility
$\chi(\bq,\omega)$ in both the energy and momentum spaces[1-24]. 
Using these data one can try to understand what is
the relation between the superconducting and magnetic properties of
high-T$_c$ superconductors. A nontrivial feature 
that have attracted a lot of attention is the so-called resonance peak 
appearing in the superconducting state and  seems to be directly 
related to the formation of the superconducting state[1-12].

Here, we will focus on the completely different feature 
of Im$\chi(\omega, {\bf q})$, namely on the off-resonance spectrum.
Substantial interest has been recently devoted to that 
contribution as incommensurate peaks have been observed away from 
the resonance peak in YBCO$_{6.6}$\cite{incdai,Mook}. However, 
it was  observed  so far in limited doping, energy and temperature 
ranges. Generally, in the 
{\it normal state} $\chi(\omega, {\bf q})$ is peaked  at the commensurate 
wavevector $(\pi,\pi)$. This 
contribution is then simply characterized by a q-width in momentum space, 
$\Delta q$ (HWHM). 

Considering the  neutron scattering data in  
YBCO for oxygen concentrations between $ x= 0.45 - 0.97$ with 
respective $T_c$ up to 93 K, we find a surprisingly simple {\em linear} 
relation between 
 superconducting transition temperature $T_c$ and HWHM $\Delta q$
 for the whole doping range:
\beqa
T_c = \hbar v^* {\Delta q}, \  \  \ \hbar v^* = 35 {\rm meV \AA}
\label{v*}
\eeqa
This observation is based on  analysis of the data and we used no theory 
assumptions in extracting the velocity $v^*$ from the data. The left-hand 
side of the above equation has dimension of energy, $\Delta q$ has an
inverse distance dimension, hence the coefficient relating them should
have dimension of {\em velocity}. A priori, it is not clear that this 
relation implies the existence of the mode with such a velocity. 
We believe it does: the magnetic properties of the YBCO are described by two   
velocities $v_{SW}$ and $v^*$.

Below, we explain how the Eq.(\ref{v*}) is obtained. The spin susceptibility  
in the metallic state of YBCO is experimentally found to have a maximum at any 
energy
at the \em{commensurate} \rm in-plane wavevector 
$q_{AF}=({h\over2},{k\over2})$ (with $h,k$ even integer), referred 
to as $(\pi,\pi)$ [1-12,16-24]. This generic rule is 
found to be violated in two   cases. 

First, in the underdoped  YBCO$_{6.6}$, 
Dai et al\cite{incdai} reported low temperature q-scans at $\hbar\omega$= 
24 meV which display well-defined double peaks\cite{oldinc}. 
Recent measurements with improved q-resolution confirm this
observation \cite{Mook}. However, this behavior is mostly observed 
at temperatures below $T_C$. In the normal state, a broad commensurate peak 
is restored in the same sample (unambiguously above 75 K)\cite{dai}.

The other case where the spin susceptibility was not found maximum 
at $(\pi,\pi)$ is above $\sim$ 50 meV in the weakly doped 
YBCO$_{6.5}$\cite{prb97}. Dispersive quasi-magnons behavior is observed
in this high energy range. Most likely, this is reminiscent of 
spin-waves observed in the undoped AF parent compound YBCO$_6$. 

Therefore, concentrating on the low energy spin excitations (below 
50 meV), \imki\ is characterized  in the {\it normal state }
by a broad maximum at the commensurate wavevector. This justifies an 
analysis in terms of a single peak centered 
around $q_{AF}$. However, the shape in q-space is systematically 
found to be sharper than a Lorentzian shape usually assumed to describe
such a disordered magnetic system. 
The neutron scattering function is then empirically found to be well 
accounted for by a Gaussian line-shape\cite{rossat,hoechst,tranquada} 
such as

\begin{equation}
S(Q,\omega) = I_{max}(\omega) \exp \Big( - \log 2 
{\frac{(q - q_{AF})^2}{\Delta_q^2(\omega)} } \Big)
\label{gaus}
\end{equation}

where $\Delta q(\omega)$ is the half width at half maximum (HWHM).
In principle, $\Delta q(\omega)$ is an increasing function of energy. 
However, a rather weak energy dependence is found for $\Delta q(\omega)$ 
with only a slight increase with the energy\cite{rossat,tranquada,note_qw}. 
Furthermore, this energy dependence becomes less pronounced for 
the higher doping range. 

The situation is even more subtle 
for $x \ge$ 0.6 as \imki\ is characterized by two distinct 
(although inter-related) contributions: one occurs exclusively in 
the superconducting state, the resonance peak, the second 
one appears in both states and is characterized by a broad peak
(around $\sim$ 30 meV). They mainly differ by their energy dependences
as the resonance peak is basically resolution limited in 
energy\cite{rossat91,tony1,hoechst,dai,epl,tony3}.
 With increasing doping, the off-resonance spectrum is 
continuously reduced (becoming too weak to be measured in the overdoped 
regime YBCO$_7$\cite{tony1,lpr,revue}) whereas the resonant peak 
becomes the major part of the spectrum\cite{revue}. The recent 
incommensurate peaks measured    below the resonance 
peak in YBCO$_{6.6}$\cite{incdai,Mook} confirms the existence of 
two contributions\cite{comment}
as the low energy incommensurate  excitations cannot belong 
to the same excitation as the commensurate resonance peak. 

At each doping, the  peak intensity at the
resonance energy is characterized by a striking temperature dependence 
either resembling to an order parameter-like dependence for the 
higher doping range (x$>$0.9) \cite{rossat,mook93,hoechst} or just
displaying a marked kink at $T_C$ \cite{rossat94,dai,epl,tonynew}.
Therefore, this mode is a novel signature of the superconducting 
state in the cuprates, and most likely is due to electron-hole pair 
production across the superconducting energy gap \cite{tony1,revue}.
In contrast, a much smoother temperature dependence is observed 
for the off-resonance spectrum\cite{epl,revue}. 
This "normal" contribution has not received much attention so far. 
However, the  knowledge  of the non resonant peak in the normal state
is important and is crucial for some proposed mechanisms for 
the high-$T_C$ superconductivity based on antiferromagnetism, e.g. \cite{pines}.

The resonance peak is related to smaller q-values (and hence larger 
real space distance) as  $\Delta q(\omega)$ exhibits a minimum at 
the energy of the resonance peak\cite{sympo,hoechst,epl}.
Furthermore, its q-width remains almost
constant whatever the doping, $\Delta q^{reso} = 0.11 \pm 0.02$ 
\AA$^{-1}$\cite{sympo}. Recent data \cite{dai,epl,tonynew} agree with 
that conclusion. Applying the simple relation $\xi=1/\Delta q$, it 
yields a characteristic length for the resonance peak, $\xi \simeq 9$ \AA\,
which might be related to the superconducting coherence length as 
the resonance peak is intimately linked to the high-$T_C$ superconductivity.
In contrast, the "normal" contribution is characterized 
by a doping dependent q-width which in terms of the Nearly AF Liquid
approach\cite{pines} would yield surprisingly small correlation length 
${{\xi}\over{a}} \simeq 1-2$ (for x$\geq$0.6).

Moreover, in all inelastic neutron scattering experiments (see e.g.
\cite{rossat,lpr,tranquada,prb91}),
the q-width is found temperature independent at any doping.
This finding is especially clear for x$\simeq$0.5 where the 
q-width at low temperature is small enough to increase upon heating.
But, in contrast, no evolution is seen within error bars (about 10 \%)
up to room temperature\cite{prb91}. For larger doping, the low 
temperature q-width is already large and the AF intensity vanishes 
without any sign of q-broadening when increasing temperature\cite{rossat}.
Therefore, these q-widths might be related to new objects essentially 
dependent on the doping level. 

To emphasize the precise value of the q-width, we have summarized in 
Table \ref{qwidth} the neutron data obtained over the last decade by 
few different groups. We consider only the low energy results for 
each oxygen content, where $\Delta q$ is weakly energy 
dependent. The energy range of interest is indicated in Table 
\ref{qwidth}. 
The $\Delta q$ value reported here has been mostly taken along the 
[110] reciprocal direction. Other data have been also taken along the
[310] reciprocal direction\cite{incdai,prb97,tonynew} which basically 
agree with the hypothesis of an isotropic q-width.

$\Delta q$  versus the oxygen content displays a double 
plateau shape\cite{sympo} which reminds the standard $x$ dependence 
of $T_C$ in YBCO. For the 90-K phase, $\Delta q^{HWHM} = 0.22$ 
\AA$^{-1}$ yielding a very short AF correlation length within 
${\rm CuO_2}$\ planes ${{\xi}\over{a}} \simeq 1.1$.

\begin{figure}
\epsfxsize=3in
\centerline{\epsfbox{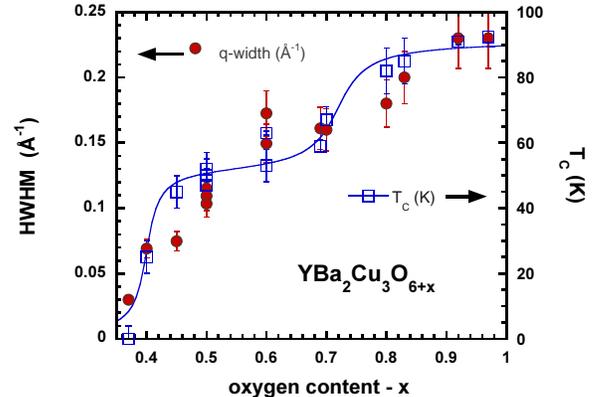}}
\caption[]{ Q-width (closed circles) and $T_C$ (open squares) versus 
oxygen content. }
\label{DqTcvsx}
\end{figure}

\begin{figure}
\epsfxsize=3in
\centerline{\epsfbox{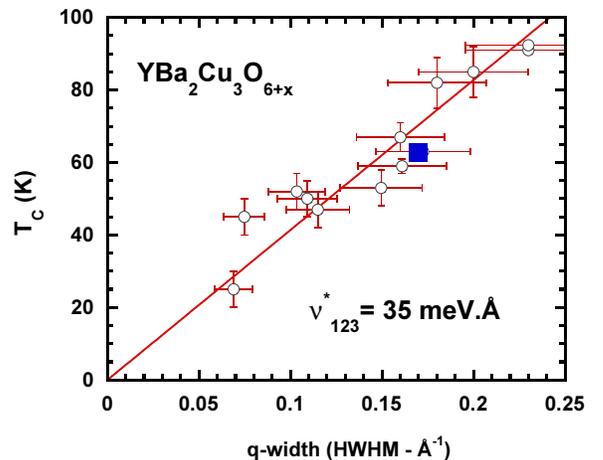}}
\caption[]{ Superconducting transition versus q-width\cite{lorentz}. 
    The full square corresponds to locus of the maximum of 
    incommensurate magnetic excitations $\delta$ recently
    reported below $T_C$\cite{Mook}.}
\label{TCvsDelta}
\end{figure}

Summarizing the data in this Table, we plot both $T_c(x)$ and 
$\Delta q(x)$ in Fig. \ref{DqTcvsx}, and find the linear 
relation between $T_c$ and $\Delta q$ (Fig. \ref{TCvsDelta}) 
in the whole oxygen doping range, Eq. (\ref{v*}), where $T_c$ 
is the respective superconducting 
transition temperature at a given oxygen concentration $x$ and 
$\Delta q$ is the corresponding {\em  half-width}  of the peak at
$(\pi,\pi)$ in $\chi"({\bf q},\omega)$. The velocity 
$\hbar v^* = 35$  meV.\AA\ is about a factor of two larger than 
the equivalent velocity in LaSrCuO, $\hbar v^*_{214} = 20$ meV.\AA, 
inferred from $T_c$ vs $\delta$ plot (Fig. \ref{Yamadaplot}), see below.  
The Eq.(\ref{v*}) does imply that the magnetic correlations, as 
measured by $\chi(\bq, \omega)$, and superconducting transition 
are closely related.

The recent incommensurate splitting $\delta$ of the peak at 
$(1/2+\delta,1/2)$ =$(1/2,1/2+\delta)$, observed 
 by Dai {\it et al} \cite{incdai} and subsequently by Mook 
{\it et al} \cite{Mook} in YBCO$_{6.6}$ have been included in 
Fig. \ref{TCvsDelta} (full square). Interestingly, the incommensuration 
$\delta$ by Mook {\it et al}\  fall on the same linear plot.

First, we  can make few comments on 
what one can extract from such a simple 
relationship as Eq. (\ref{v*}) regardless of the particular mechanism 
responsible for $v^*$:

a) Proportionality between the width of the peak and the critical 
temperature implies that there is a characteristic  velocity $\hbar v^*$
which is the same (within experimental resolution) for a wide range of 
oxygen doping in YBCO. 

b) Velocity $\hbar v^*$ is two orders of magnitude smaller than typical 
Fermi velocity $\hbar v_F \sim 1$ eV.\AA\  in these compounds 
\cite{Loeser}. This perhaps is not surprising as we are considering 
the magnetic response where localized Cu spins provide the main 
contribution. 

c) More importantly, $v^* << v_{SW}$ is about an order of 
magnitude smaller than the spin wave velocity, $\hbar v_{SW} \simeq 0.65$  
eV.\AA\cite{shamoto}, of the parent compound but also much smaller than 
the spin velocity in the metallic state $\hbar v_{spin} \simeq 0.42$ 
eV.\AA\cite{prb97}. This is a nontrivial fact. It may suggest that some 
magnetic soft mode is present. We do not have a  model to 
explain the data presently. On the other hand  on general 
grounds for any approach, based on the simple spin-wave theory, one 
would expect the typical spin-wave velocity to characterize the width 
in $\chi''(\bq, \omega)$. 

d) If, as we are proposing, the characteristic velocity $\hbar v^*$ does 
correspond to some  propagating or diffusive mode than there should be 
a way to directly observe it in other experiments.

Now we would like to discuss the possible origin of $\hbar v^*$.   
It is  likely caused by some  phase fluctuation mode associated with the slow 
motion of density excitations. These could be caused by ''stripe'' fluctuations.
  Recent tunneling and photoemission studies indicate that the gap in the 
SC state increases as Tc decreases on underdoping \cite{underdoping}.  
The  $T_c$  would then been determined by 
phase fluctuations, as emphasized by Emery and Kivelson \cite{EK}. Recently, 
based on 
phase fluctuations
 model, the similarly small velocity ($60 meV \AA$ for LaSrCuO) was obtained by 
Casto Neto \cite{antonio}. It 
is therefore natural that  phase mode velocity will determine the 
superconducting temperature.
 
  The existence of the second magnetic velocity $v^*$ we interpret as a 
closeness 
to the quantum critical point (QCP), controlled by some density instability with 
strong coupling to the spin channel. The antiferromagnetic insulating 
compound  with characteristic $v_{SW}$ determines the critical point  at zero 
doping, where the transition into 3D antiferromagnetic state occurs at finite 
temperature. The 
second critical point  with characteristic $v^*$  along the doping axis  is  
close to the optimally doped compound 
and might be    determined by some density wave or ''stripe'' instability. Based 
on the neutron scattering data for LaSrCuO system this point was emphasized by 
Aeppli et.al. \cite{QCP214}. 

Within the simple model, say t-J, the  energy scales are set by $t$, which 
determines $v_F$ and by $J$, determining $v_{SW}$. Hence one generally would not 
expect any excitations in this model with $v^*$. One possibility to generate a 
new energy scale in the problem is to allow  some
(microscopic) inhomogeneities.  Phase separation into hole-rich and 
antiferromagnetic regions  with  fluctuations of the boundaries between regions
   will occur   with some soft velocity that might be 
related to $v^*$.

Finally,  we would like to relate the above discussion to the other well 
studied system: LaSrCuO.  Inelastic neutron scattering data by Yamada 
{\it et al} \cite{Yamada} on LaSrCuO (La214) compounds show the existence 
of the incommensurate peaks at $(\pi \pm \delta,\pi)$ and $(\pi, \pi
\pm \delta)$ \cite{Yamada}.  Plotted vs $\delta$,  $T_c(\delta)$ 
was found to be a linear function of $\delta$ in the wide range of 
Sr doping, see Fig. 3, as appear in \cite{Yamada}. Using the same reasoning 
as for Eq.(\ref{v*}) from the  data \cite{Yamada}  we find the 
characteristic {\em velocity}, 
\begin{equation}
T_c = \hbar v^*_{214} \delta ,  \ \ \hbar v^*_{214} = 20 meV.\AA
\end{equation}
Thus inferred velocity is  much smaller that the Fermi 
velocity on La214 $\hbar v_F \sim  1-0.5$ eV.\AA\ and smaller than the 
measured spin wave velocity $\hbar v_{SW} \sim 0.85$ eV.\AA\cite{aeppli}. 
We should again emphasize that the linearity $T_c$ vs $\delta$ is an 
experimental 
fact. The coefficient relating energy scale $T_c$ to the inverse length scale
$\delta$ has a dimension of  velocity.   
This result is similar to the small velocity we find for the 
$\Delta q$ vs $T_c$ plots in YBCO, except in YBCO the velocity $v^*$ is 
about a factor of two larger than in LaSrCuO.  

\begin{figure}
\epsfxsize=3in
\centerline{\epsfbox{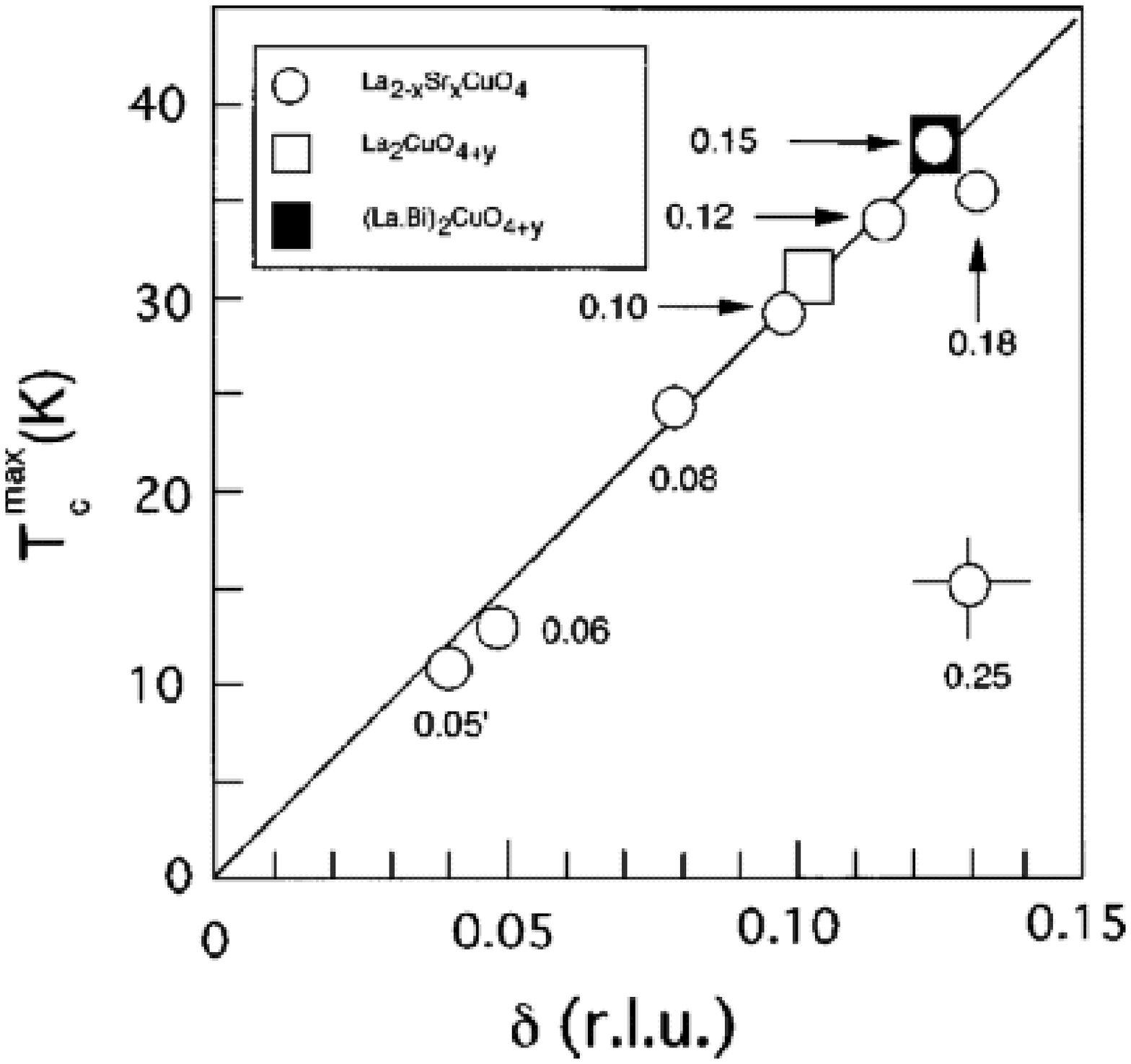}}
\caption[]{ Linear relation between the incommensurate peak splitting 
$\delta$ and $T_c$ for the LaSrCuO superconductors. The linear slope of 
the curve means that $T_c = \hbar v^*_{214} \delta$. We find $v^*_{214} 
= 20  meV$.\AA. Note that the linear dependence is violated in the 
overdoped regime.  The figure is taken from 
Yamada et.al. (Fig. (10)) \cite{Yamada}}
\label{Yamadaplot}
\end{figure}

In conclusion, we find that the body of neutron scattering data on YBCO for a 
wide range of oxygen doping allows the simple linear relation between the width 
of the ``normal'', i.e. out of resonance peak, $\delta q$ and corresponding 
$T_c$ of the sample. Thus inferred velocity $\hbar v^* \simeq$ 35 meV \AA\ is 
anomalously small compared to known spin wave and Fermi velocities for these 
compounds.  We suggest that this velocity indicates the existence of some new  
mode in these materials and this mode is closely related to the formation of 
superconducting state. 
 

We are grateful to G. Aeppli, L.P. Regnault, R. Silver, Y. Sidis and  J. 
Tranquada  for 
the 
useful discussions. This work was supported by the US DOE.


 \newpage

\widetext

\begin{table}[h]
\begin{tabular}{c c c c c c c c c c c c}
\hline
x & 0.4 & 0.45 & 0.5 & 0.5& 0.5 &0.6 & 0.6\\
$T_C$ (K) & 25 & 45 & 47& 50 & 52& 53 & 63\\
Energy range (meV)& 3-15& 3-15 & 4-15& 2-15 & 4-15 & 6-15 & 24$^*$ \\
\hline
$\Delta_{q}^{mes}$ (\AA$^{-1}$) & 0.075 & 0.08 & 0.12 & 0.115 & 0.115 & 0.14 
& 0.18 \\
$\Delta_{q}^{resol}$ (\AA$^{-1}$)& 0.023 & 0.023 & 0.023 & 0.05 & 0.05 & 
0.035 & 0.07\\
$\Delta_{q}$ (\AA$^{-1}$) & 0.07 & 0.075 & 0.115 & 0.11 & 0.1 & 0.14 & 
0.17\\
\hline
Refs.& \cite{chou} & \cite{chou} &  \cite{rossat,lpr}& 
\cite{prb97,tonynew}& \cite{prb91} &\cite{tranquada,sternlieb} & \cite{dai} 
\\
\end{tabular}

\vspace {0.2 cm}

\begin{tabular}{c c c c c c c c c c c c}
\hline
x &0.69& 0.7& 0.8 & 0.83 & 0.92&  0.97\\
$T_C$ (K)  & 59 & 67 & 82& 85 & 91 & 92.4\\
Energy range (meV) & 15-20 & 15-25 & 20-25 & 15-25 & 28-38 &  33-37 \\
\hline
$\Delta_{q}^{mes}$ (\AA$^{-1}$) & 0.17 & 0.17 & 0.21 & 0.23 & 0.25 & 0.25\\
$\Delta_{q}^{resol}$ (\AA$^{-1}$) & 0.06 & 0.07 & 0.09 & 0.11 & 0.11 & 0.11 
\\
$\Delta_{q}$ (\AA$^{-1}$) & 0.16 & 0.16 & 0.18 & 0.2 & 0.23 & 0.23\\
\hline
Refs.& \cite{rossat_old} & \cite{tonynew} & \cite{epl} & \cite{rossat94} 
\cite{rossat91,lpr} & \cite{hoechst}\\
\end{tabular}
\caption[table] { \label{qwidth}
Q-widths (HWHM) of the AF intensity as a function of the oxygen 
content at energies corresponding to the non-resonant contribution. 
Q-widths have been mostly determined along the [110] direction (see 
text). $\Delta_{q}^{resol}$  denotes the HWHM of the Gaussian 
resolution of the spectrometer. The intrinsic q-widths have been 
then obtained after deconvolution from the spectrometer linewidth 
assuming a Gaussian shape wavevector dependence for \imki\  (Eq. 
\ref{gaus}). Energy range from where the q-width has been taken
 is given ($^*$ only reported value).
Fully oxidized YBCO$_7$ is not quoted in this table  as the
normal AF contribution is not detectable in this overdoped regime 
\cite{lpr,tony1,revue}. 
} 
\label{table}
\end{table}


\begin{references}

\bibitem{rossat91} J. Rossat-Mignod, L.P. Regnault, C. Vettier, P. Bourges,
P. Burlet, J. Bossy, J.Y. Henry and G. Lapertot,  Physica C, {\bf 185-189},
  86, (1991).

\bibitem{rossat} 
J. Rossat-Mignod, L.P. Regnault, P. Bourges, C. Vettier, P. Burlet 
 and J.Y. Henry : in \it{Selected Topics in Superconductivity} \rm 
Frontiers in Solid State Sciences Vol 1., eds L.C. Gupta and M.S. Multani
 (World Scientific 1993), 265.

\bibitem{mook93} H.A. Mook, M. Yethiraj, G. Aeppli, T.E. Mason 
and T. Armstrong : Phys. Rev. Lett., {\bf 70}, 3490, (1993).

\bibitem{rossat94} 
 J. Rossat-Mignod, P. Bourges, F. Onufrieva, L.P. Regnault,  P. Burlet, C. 
Vettier, and J.Y. Henry, Physica B,{\bf 199\&200}, 281, (1994).

\bibitem{tony1}  H. F. Fong, B. Keimer, P.W. Anderson, 
D.   Reznik, F. Dogan, and I.A. Aksay, {\rm Phys. Rev. Lett.}, 
{\bf 75}, 316 (1995).

\bibitem{sympo}  P. Bourges, L.P. Regnault, J.Y. Henry,
C. Vettier, Y. Sidis, and  P. Burlet, {\rm Physica B}, {\bf 215}, 30, 1995).

\bibitem{lpr} L.P. Regnault, P. Bourges,  P. Burlet,  J.Y. Henry, 
J. Rossat-Mignod, Y. Sidis, and C. Vettier, {\rm Physica C}, 
{\bf 235-240}, 59, (1994); {\rm Physica B}, {\bf 213\&214}, 48, (1995).

\bibitem{hoechst}  P. Bourges, L.P. Regnault, Y. Sidis, and C. Vettier, 
{\rm Phys. Rev. B}, {\bf 53}, 876, (1996).

\bibitem{dai} P. Dai, M. Yethiraj, H.A. Mook, T.B. Lindemer, 
   and  F. Dogan, Phys. Rev. Lett. {\bf 77},  5425, (1996).

\bibitem{epl}  P. Bourges, L.P. Regnault, Y. Sidis, J. Bossy, P. Burlet,
 C. Vettier, J.Y. Henry, and M. Couach, Europhysics Lett. 
{\bf 38}, 313 (1997).

\bibitem{tony3}  H.F. Fong, B. Keimer, F. Dogan, and I.A. Aksay, Phys.
Rev. Lett. {\bf 78}, 713 (1997).

\bibitem{revue} P. Bourges  { in \it The gap Symmetry and Fluctuations in 
High Temperature Superconductors}  Edited by J. Bok, G, deutscher, 
D. Pavuna and S.A. Wolf. (Plenum Press, 1998); L.P. Regnault, P. Bourges, 
and P. Burlet in {\it Neutron Scattering in Layered Copper-Oxide 
Superconductors }, Edited by A. Furrer, 85 (Kluwer, Amsterdam, 1998).

\bibitem{incdai} P. Dai, H.A. Mook, and  F. Dogan, Phys. Rev. Lett. 
{\bf 80},  1738, (1998) (cond-mat/9707112).

\bibitem{Mook} H.A.  Mook, P. Dai, S.M. Hayden, G. Aeppli, T.G. Perring and F. 
Dogan, Nature, {\bf 395}, 580, (1998). See also ISIS report at 
http://www.isis.rl.ac.uk/ISIS98/feat11.htm.

\bibitem{oldinc} Incommensurability has been previously inferred 
in YBCO$_{6.6}$ from such flat-topped shape profiles \cite{tranquada,sternlieb}.

\bibitem{tranquada} J.M. Tranquada, P.M. Gehring, G. Shirane, S. Shamoto 
          and M. Sato, Phys. Rev. B, {\bf 46}, 5561 (1992).

\bibitem{sternlieb} B.J. Sternlieb, J.M. Tranquada, G. Shirane,  
          M. Sato, and S. Shamoto, Phys. Rev. B, {\bf 50}, 12915 (1994).

\bibitem{prb97}   P. Bourges, H.F. Fong, L.P. Regnault, J. Bossy, 
C. Vettier, D.L. Milius, I.A. Aksay, and B. Keimer, Phys. 
Rev. B, {\bf 56}, R11439, (1997) (cond-mat/9704073).

\bibitem{note_qw} Recent measurements for $x=0.5$ and  $x=0.7$ 
             confirm this trend\cite{prb97,tonynew}.

\bibitem{comment}  P. Bourges, and L.P. Regnault, Phys. Rev. Lett.,
  {\bf 80}, 1793, (1998); P. Dai, M. Yethiraj, H.A. Mook, 
   T.B. Lindemer, and  F. Dogan, Phys. Rev. Lett. {\bf 80}, 1794, (1998).

\bibitem{tonynew}  H.F. Fong, P. Bourges,  B. Keimer, L.P. Regnault, 
J. Bossy,  A.S. Ivanov, D.L. Milius,  and I.A. Aksay, in 
preparation,

\bibitem{prb91} P. Bourges, P.M. Gehring, B. Hennion, A.H. Moudden, 
            J.M. Tranquada, G. Shirane, S. Shamoto, and M. Sato, 
            Phys. Rev. B, {\bf 43}, 8690, (1991).  

\bibitem{chou} H. Chou, J.M. Tranquada, G. Shirane, S. Shamoto 
          and M. Sato, Phys. Rev. B, {\bf 43}, 5554 (1991).

\bibitem{rossat_old} J. Rossat-Mignod, L.P. Regnault, C. Vettier, 
P. Burlet,  J.Y. Henry and G. Lapertot, Physica B, {\bf 169}, 58 (1991).

\bibitem{pines} D. Pines, Z. Phys. B, {\bf 103 }, 129 (1997).

\bibitem{lorentz} It should be noted that a fit of the AF fluctuations 
by a Lorentzian shape reduce the q-width by about 15 \%.
Then this would  affect the velocity $v^*$ by the same amount. We are interested 
in the peak width and not in the tails away from the peak.

\bibitem{shamoto} S. Shamoto, M. Sato, J.M. Tranquada, B. Sternlieb, and 
G. Shirane, Phys. Rev. B, {\bf 48}, 13817 (1993).

\bibitem{Loeser} The Fermi velocity $v_F \simeq 1$ eV.\AA\  can be 
inferred from photoemission data of A. G. Loeser {\it et al.}, Science, 
{\bf 273}, 325 (1996), and of H. Ding,  {\it et al.}, Nature, 
{\bf 382}, 51, (1996).  
This estimate is consistent with the transport measurements,  see 
for example,  K. Krishana, J. Harris and P. Ong, Phys. Rev. Lett., 
{\bf 75}, 3529, (1995).


\bibitem{aeppli} G. Aeppli, S.M. Hayden, H.A. Mook, Z. Fisk,  S-W. Cheong,
D. Rytz, J.P. Remeika, G.P. Espinosa, and A.S. Cooper, Phys. Rev. Lett., 
{\bf 62}, 2052 (1989).

\bibitem{underdoping} M. Oda et al., Physica C 281, 135 (1997); C. Renner et 
al., Phys. Rev. Lett. {\bf 80}, 149 (1998); N. Miyakawa et al., Phys. Rev. Lett. 
{\bf 80}, 157 (1998); J. M. Harris et al., Phys. Rev. {\bf B 54}, R15665 (1996); 
H. Ding et al., Nature {\bf 382}, 51 (1996); D. N. Basov et al., Phys. Rev. 
Lett. {\bf 77}, 4090 (1996).



\bibitem{EK} V.J. Emery and S.A. Kivelson, Physica, {\bf 209C}, 597, (1993); 
V.J. Emery and S.A. Kivelson, Nature {\bf 374}, 434, (1995).  

\bibitem{antonio}   A. 
Castro Neto, Phys. Rev. Lett., {\bf 78 }, 3931, (1997).

\bibitem{QCP214} G. Aeppli, et. al., Science, {\bf 278}, 1432, (1997).



\bibitem{Yamada} K. Yamada, et.al., Phys. Rev. {\bf B 57}, 6165, (1998)

 \end{references}
\end{document}